\documentstyle[preprint,prb,aps]{revtex}
\begin{document}
\draft

\title{Variational and DMRG studies of the frustrated
antiferromagnetic Heisenberg $\mathbf{S=1}$ quantum spin chain}

\author{A. Kolezhuk}
\address{Institut f\"{u}r Theoretische Physik,
Universit\"{a}t Hannover, 30167 Hannover, Germany}
\address{\protect\cite{perm}Institute of Magnetism, Natl. Acad. Sci. of 
Ukraine, 252142 Kiev, Ukraine }
\author{R. Roth and U. Schollw\"{o}ck}
\address{Sektion Physik, Ludwig-Maximilians-Univ.\ M\"{u}nchen,
Theresienstr.\ 37, 80333 Munich, Germany}

\date{September 4, 1996}

\maketitle
\begin{abstract}
In this paper we study a frustrated antiferromagnetic isotropic
Heisenberg $S=1$ quantum spin chain
\[
H = \sum_{i} {\mathbf S}_{i} {\mathbf S}_{i+1} +
\alpha \sum_{i} {\mathbf S}_{i} {\mathbf S}_{i+2}, 
\]
using a variational
ansatz starting from valence bond states and the Density Matrix
Renormalization Group. We find both methods to give results in
very good qualitative and good quantitative agreement, which
clarify the phase diagram as follows: At $\alpha_D=0.284(1)$,
there is a disorder point of the second kind, marking the onset
of incommensurate spin-spin correlations in the chain. 
At $\alpha_L=0.3725(25)$
there is a Lifshitz point, at which the excitation spectrum is found
to develop a particular doubly degenerate structure. These points
are the quantum remnants of the transition from antiferromagnetic
to spiral order in the classical frustrated chain. At
$\alpha_T=0.7444(6)$ there is a first order phase transition
from an Affleck-Kennedy-Lieb-Tasaki (AKLT) phase characterized by 
non-vanishing string order
to a phase which can be understood as a next-nearest
neighbor generalization of the AKLT model. 
At the transition,
the string order parameter shows (to numerical precision) a
discontinuous jump of 0.085 to zero; the correlation length and the gap 
are both finite
at the transition.
The problem of edge states in open frustrated 
chains is discussed at length.
\end{abstract}

\narrowtext

\section{Introduction}

Over the past years, frustrated quantum mechanical systems have
met with considerable interest.\cite{Diep 94} Research is concentrating
on geometrically frustrated systems, where the lattice
geometry introduces competing interactions, and on systems, where
frustration is directly introduced by interaction, e.g.\ the additional
presence of next-nearest neighbor interactions. Interest in
these systems has twofold motivation. There are
experimental systems which show geometrical frustration (quasi 1D
hexagonal insulators such as CsNiCl$_3$ or effective Kagom\'{e}
lattices such as SrCr$_{8-x}$Ga$_{4+x}$O$_{19}$); there are also
systems frustrated by interaction: A well-known motivation is
given by high-$T_{c}$ superconductors, which exhibit an
antiferromagnetic phase. It has been shown early on that doping
of these compounds may be mapped to frustrated spin
models.\cite{Inui 88} Another prominent example is CuGeO$_3$, a
one-dimensional S=1/2 antiferromagnet with strong competing interactions.
\cite{Castilla+95}
 On the other hand, there is purely
theoretical motivation to study these systems. Taking the
classical limit, frustration may introduce rather complex forms of
order. Quantum
fluctuations will destroy long-range order which may be present in the
classical system, if the dimension of the system is low. For
one-dimensional quantum systems, 
order will in general be destroyed even at
zero temperature.\cite{Mermin 66} It is therefore of interest to
study which remnants of the classical system survive in the
quantum system, and whether there are purely quantum phenomena present.
About the simplest frustrated system conceivable is the frustrated
Heisenberg isotropic quantum spin chain with antiferromagnetic
interactions between nearest and next-nearest neighbors.
{}From the  
well-established\cite{White 92,White 93a,Golinelli 94} 
Haldane conjecture\cite{Haldane 83a,Haldane 83b} 
it is known that in the limit of a vanishing
next-nearest-neighbor interaction there is a fundamental
difference between half-integer and integer spin chains. The
unfrustrated half-integer spin chain is characterized by almost
long-range antiferromagnetic order, with power-law correlations
and a critical spectrum $\omega=c(k-\pi)$ at low energies. The
integer spin chain has only short-ranged antiferromagnetic order,
exponential correlations and a gapped spectrum
$\omega=\{c^2(k-\pi)^2+\Delta^2\}^{1/2}$, where $\Delta$ is the
Haldane gap. We may thus expect considerably different behavior
also in the frustrated chains.

The case of half-integer spin chains has been extensively studied
and is by now well understood.\cite{White 96} The case of
frustrated isotropic next-nearest neighbor integer spin chains 
\begin{equation}
H = \sum_{i} {\mathbf S}_{i} {\mathbf S}_{i+1} +
\alpha \sum_{i} {\mathbf S}_{i} {\mathbf S}_{i+2}
\label{ham}
\end{equation}
has also
attracted considerable interest. Several scenarios and analytical
and numerical studies have been proposed, in particular for
$S=1$. Numerical studies\cite{Tonegawa 92,Harada 93} seem to
indicate that there is no phase transition for any value of
frustration. Field theoretical studies\cite{Rao 94,Allen 95}
predict that there is always a gap for any value of frustration
They indicate a doubly degenerate excitation spectrum beyond 
$\alpha\approx 0.4.$\cite{Allen 95} 
On the other hand, it was claimed
recently\cite{Pati 96a,Pati 96b} that there is an almost gapless
point (to numerical precision) for $\alpha=0.73(1)$.
The situation is thus obscure; there
is no agreement whether there is a phase transition in the chain and
if so, of which order.

In this work we study the phase diagram of a 
frustrated antiferromagnetic 
isotropic Heisenberg
quantum spin chain (\ref{ham})
with $S=1$ at $T=0$. 
We start with a variational approach
based on the 
Affleck-Kennedy-Lieb-Tasaki (AKLT)\cite{Affleck 87,Affleck 88} 
model to give analytical predictions. Basically,
two ground states are compared to each other, one being the
conventional AKLT model, the other an AKLT model which links
next-nearest neighbors by singlet bonds (Fig.\ \ref{fig:aklt}).
For further refinement, an
ansatz interpolating between these cases is constructed and
minimized.
Elementary excitations are calculated in a 
``crackion''\cite{FathSolyom93,NeuMik96}
picture.
On the other hand, we
use the Density Matrix Renormalization Group 
(DMRG)\cite{White 92,White 93b} to obtain
quantitative results and to check the variational approach. In
our DMRG calculations, we typically use $M=250$ block states
in chains up to $L=380$ sites. This by far
exceeds previous calculations\cite{Pati 96a,Pati 96b} in
precision. We present calculations of important quantities not considered
beforehand and analyze the excitation spectrum carefully.
The use of a prediction mechanism\cite{Scholl 96z} to accelerate the
convergence of the exact diagonalization inherent in the DMRG
allows us to go up to about $M=400$ in a $S=1$ system on a
PentiumPro based personal computer while retaining reasonable
computing speed. The structure of the paper is as follows: in
section II, we briefly summarize our findings; in section III, we
discuss the associated physical scenario in more detail. Sections
IV and V present the analytical and numerical calculations.

\section{Summary of Results}

We find that there is an AKLT (Haldane) phase for
$\alpha_T<0.7444(6)$ with a disorder point of the second
kind\cite{Stephenson 69} at $\alpha_D=0.284(1)$ (see Figs.\
\ref{fig:correl_disorder} and \ref{fig:real_disorder}) and a
Lifshitz point at $\alpha_L=0.3725(25)$ (see Fig.\
\ref{fig:sq_disorder}). The disorder point and the associated
Lifshitz point can be understood as the quantum remnants of the
phase transition from antiferromagnetic to spiral order in the
classical frustrated chain. The AKLT phase can be understood in
terms of the conventional $S=1$ AKLT model (Fig.\
\ref{fig:aklt}); the associated string order parameter\cite{Nijs
89} is nonzero (Fig.\ \ref{fig:sop}), corresponding to hidden
order linked to the breaking of a discrete $Z_2 \times Z_2$
symmetry in the chain.\cite{Kennedy 92} The string order
parameter shows a maximum of $0.4397(1)$ at $\alpha=0.3375(25)$
close to the disorder point, and disappears with a discontinuous
jump of about $0.085$ at $\alpha_T=0.7444(6)$. The gap rises
monotonically to a maximum $\Delta=0.82(1)$ at $\alpha=0.40(1)$,
then drops monotonically (Fig.\ \ref{fig:gap}). Just before the
disorder point, the correlation length drops steeply (in all
probability, with infinite slope close to the disorder point) to
reach a minimum of $\xi_D=1.20(2)$ exactly at the disorder point,
and rises slowly and monotonously beyond (Fig.\
\ref{fig:correl_disorder}). The real-space correlations are
purely antiferromagnetic below the disorder point, but
incommensurate above (Fig.\ \ref{fig:real_disorder}). The change
in the wave vector $q$ is discontinuous at the disorder point
(Fig.\ \ref{fig:q_disorder}). The bulk excitations for
$\alpha>\alpha_L$ are characterized by a pairwise degeneracy of
states of equal spin, but different parity. This degeneracy is
not present for $\alpha<\alpha_L$. The variational approach
predicts the Lifshitz point at $\alpha_{L}^{var}=0.32$ and a gap
maximum at $\alpha=0.38$ with $\Delta^{var}(0.38)=0.97$, and the
presence of the AKLT phase up to $\alpha_{T}^{var}=0.75$ (or
$\alpha_{T}^{var}=0.81$, depending on the approach).

Beyond $\alpha_T=0.7444(6)$, we can characterize the system by a
next-nearest neighbor generalization of the AKLT-model, taking
all odd and all even numbered spins separately and defining the
usual AKLT-model on each subchain (Fig.\ \ref{fig:aklt}).
Defining a conventional string
order parameter on each subchain, we find this string correlator
to decay exponentially on a scale much longer than the bulk correlation
length for $\alpha$ greater than about 1 (Fig.\ \ref{fig:sop12}).
As the $Z_2 \times Z_2$ symmetry is not broken in the total
chain, the string correlator is not an order parameter, but its very
slow decay shows it characterizes well the NNN-AKLT phase. The gap
increases with frustration. In the $\alpha\rightarrow\infty$
limit, it tends towards the unfrustrated value
$\Delta=0.41050(2) \alpha$ due to the chain decomposition. The
correlation length first decreases; in the large frustration
limit it tends towards twice the unfrustrated value, $\xi=12.04$,
due to the doubling of the lattice spacing in the decomposed
chains. The bulk excitations retain the pair structure mentioned above. 

At the transition point $\alpha_T=0.7444(6)$ we measure the
correlation length to saturate to $\xi_T\approx 18$ on the AKLT
side of the transition, clearly
excluding a critical or near-critical point as conjectured
before. We find the bulk excitation gap to be
$\Delta_T\approx 0.10$.  On the NNN-AKLT side the
correlation length is longer, and might be divergent. 
In both phases we can identify
close to the transition edge states that are remnants of the other
phase. 
This together with the discontinuous jump
of the string order parameter by $\approx 0.085$, about 20 percent
of its maximal value, leads us to classify this
point as a first-order transition.
This is in accordance with the predictions of the
variational approach, predicting a (too big) gap of
$\Delta_T^{var}=0.325$.

\section{Physical Scenario}

We now proceed to expound the physical scenario briefly outlined
above. The physics of the frustrated $S=1$ quantum spin chain is
entirely determined by the parameter $\alpha$. We
find two phases, namely the {\em AKLT (Haldane)} and the {\em NNN-AKLT}
phase, and three special points in the phase diagram, the disorder
point $\alpha_{D}$, the Lifshitz point $\alpha_L$  
and the transition point $\alpha_T$.

\subsection{The AKLT Phase} 

The point $\alpha=0$ corresponds to a nearest-neighbor
Heisenberg quantum spin chain. It is well established by
now\cite{White 92,White 93a,Golinelli 94} that at $\alpha=0$ the
above model has a finite correlation length of $\xi=6.03(2)$, a
Haldane gap $\Delta=0.41050(2)$, and is characterized by a
so-called string order parameter\cite{Nijs 89,Girvin 89}
\begin{equation}
{\cal O}_{\pi}^{z}(i,j) = \langle S^{z}_{i} ( \exp \sum_{k=i+1}^{j}i\pi
S^{z}_{k})S^{z}_{j} \rangle
\end{equation}
which measures the {\em hidden order} of the $S=1$ Heisenberg
chain, which is due to a broken $Z_{2}\times Z_{2}$
symmetry.\cite{Kennedy 92} Its numerical value\cite{White 93a} is
${\cal O}_\pi^z=\lim_{|i-j|\rightarrow\infty}{\cal O}_{\pi}^{z}(i,j)
= 0.374325096(2)$. 
There is a very intuitive description of this
point provided by the Affleck-Kennedy-Lieb-Tasaki
model:\cite{Affleck 87,Affleck 88} each $S=1$ spin is decomposed
into a symmetric sum of two $S=\frac{1}{2}$ spins (Fig.\
\ref{fig:aklt}). Each of the $S=\frac{1}{2}$ spins is linked to
one of the neighboring $S=\frac{1}{2}$ spins by a singlet bond.
The adequacy of this description is established by the fact that
the AKLT model shows the same hidden order characterized by a
string order parameter\cite{Nijs 89,Girvin 89} ${\cal O}_\pi^z=\frac{4}{9}$,
has a gap to the first bulk excitation, and a finite correlation
length. One of the most striking predictions of the AKLT model
verified by the Heisenberg chain is the presence of two
effectively free $S=\frac{1}{2}$ spins at the right and left end
of an open chain. This gives rise to a low-lying edge excitation triplet,
the so-called Kennedy triplet, which degenerates with the ground
state singlet in the thermodynamic limit.\cite{Kennedy 90}

We argue that for $\alpha_T<0.7444(6)$ the AKLT model provides an
adequate description and a good starting point for a variational
description of the spectrum. This is corroborated by the
observation that the string order parameter is non-zero
throughout this phase (Fig.\ \ref{fig:sop}), and peaks at
$0.4397(1)$, very close to the AKLT value of $\frac{4}{9}$. It drops to
zero discontinuously at the phase transition, and is thus an adequate
order parameter for this phase. The results obtained by the
variational approach starting from the AKLT model are in a qualitative 
agreement with the numerical findings: We observe a gap maximum
at $\alpha=0.40(1)$ with $\Delta=0.82(1)$, which is predicted at
$\alpha=0.38$ with $\Delta^{var}=0.97$ by the variational approach.
Predicted and observed gap curve are in reasonable 
agreement (Fig.\ \ref{fig:gap}).
Let
us however note at this point that we do not find the
intermediate drop of the gap Pati el al.\cite{Pati 96a,Pati 96b}
observe at $\alpha \approx 0.5$, but rather observe a monotonous
drop of the gap. For a (rather technical) explanation of the disagreement,
which can be traced to the appearance of parasitical edge states which
were not taken completely into account,
we refer to Section V.

In the region where the string order
parameter peaks, numerical and variational ground state energies agree
very well, and the general behavior of the ground state energy is well
predicted by the variational approach (Fig.\ \ref{fig:ground}).

The shift in the structure function peak away from $\pi$
(explained below) is predicted variationally at $\alpha_{L}^{var}=0.32$ 
and found 
numerically at
$\alpha=0.3725(25)$. However, in the variational approach the 
peak shifts to $q=\pm 2\pi/3$ as $\alpha$ approaches the transition point,
and then drops discontinuously to $\pm \pi/2$, 
whereas we observe numerically that
$q$ smoothly decreases from $\pi$ to $\pi/2$ with increasing $\alpha$. 
Here, the variational approach is
too simplistic. It is no use comparing correlation lengths: The
matrix product states of the AKLT model notoriously underestimate
correlation lengths. 

We conclude by remarking that we clearly
observe the Kennedy triplet numerically, as a $S_{total}=1$
end excitation with odd parity (abbreviated in the following as $1-$), 
degenerating with the $0+$
ground state. The first bulk excitation is
then given by the lowest $2+$ state. 
We have also calculated the spectrum in a chain where a
spin $\frac{1}{2}$ was added at each end, binding the free spins
and lifting the degeneracy.\cite{White 92,White 93a} Then the
ground state is given by the lowest $0-$ state, 
and the first bulk excitation by the lowest
$1+$ state. The findings of both procedures are in
excellent agreement. In an open chain, there is for $\alpha>\alpha_D$
a low-lying $1+$ edge excitation in addition to the Kennedy triplet,
which we will identify as a precursor of the NNN-AKLT phase below.

\subsection{The Disorder Point in the Frustrated Spin Chain} 

In a previous work\cite{Scholl 96a} by one of us (U.Sch.) it was
shown that the relationship between the antiferromagnetic
Heisenberg model and the AKLT model for $S=1$ can be understood
within the framework of a {\em disorder point of the second
kind}, a well-defined concept arising in classical statistical
mechanics. Let us briefly review the physical properties of such
a disorder point\cite{Stephenson 69,Garel 86} without
mathematical details; for such details, we refer to the previous
work\cite{Scholl 96a} and the references cited therein.

A disorder point can arise if a system exhibits two ordered
low-temperature phases with differently broken symmetries, e.g.\
an antiferromagnetically ordered phase and a spiral phase with a
generally incommensurate wave number $q$ and if these phases are
linked to the disordered high-temperature phase by continuous
phase transitions. It is intuitively clear that in the disordered
phase there will be
remaining short-range correlations of the type found in the
adjacent ordered phases. Moving across the phase diagram, one
expects to find a line in the disordered phase separating regions
with the two different types of correlations, a so-called
disorder line. As the correlations are short-ranged, the
correlation function peak in momentum space (in $S(q)$) starts
moving from, say, $q=\pi$ to a general $q$ not exactly at the
disorder line, but at a different line, the so-called Lifshitz
line, somewhere in the incommensurate-correlation region. 
The two lines join
in the multicritical point, where the two ordered phases and the
disordered phase meet.

The remarkable fact is that such disorder lines, throughout a
variety of substantially different physical systems, can be
classified into two types with a number of well-defined physical
properties.

Typical properties of
a {\em disorder line of the second kind} (the one found here) are:

\begin{itemize} 

\item Moving through parameter space on a path characterized by a
parameter $\gamma$ across a disorder line at $\gamma_{0}$, the
correlation length $\xi(\gamma)$ 
exhibits an infinite slope on the commensurate
side at $\gamma_{0}$. On the incommensurate side, the slope is
typically finite:
\begin{equation}
\begin{array}{lcr}
\left . \frac{d\xi}{d\gamma} \right|_{\gamma_0;C} = \infty &
\mathrm{\ \ \ \ \ \ } & 
\left . \frac{d\xi}{d\gamma} \right|_{\gamma_0;IC} < \infty 
\end{array}
\end{equation}
The correlation length may, but need not have
a local minimum at $\gamma_{0}$.

\item The real space correlation function changes from a
commensurate to an incommensurate wave number $q(\gamma)$ on the disorder
line. The function of the wave number $q(\gamma)$ has a singular
derivative at $\gamma_{0}$ on the incommensurate side 
\begin{equation}
\begin{array}{lcr}
\left . \frac{dq}{d\gamma} \right|_{\gamma_0;C} = 0 &
\mathrm{\ \ \ \ \ \ } & 
\left . \frac{dq}{d\gamma} \right|_{\gamma_0;IC} = \infty 
\end{array}
\end{equation}
and evolves as
\begin{equation}
(q(\gamma)-q(\gamma_{0}))=(\gamma-\gamma_{0})^{\sigma}
\end{equation}
for small $\gamma-\gamma_{0}$.
\item At $\gamma_{0}$, there is a {\em dimensional reduction} of
the real space correlation function. This means that comparing
the correlation function to an Ornstein-Zernike correlation
function
\begin{equation}
\langle S(0)S(x) \rangle \propto e^{-x/\xi}/r^{(d-1)/2},
\end{equation}
the underlying problem seems to have a lower dimension than the
original physical problem, i.e.\ $d<1+1$ here.

\end{itemize}

In the case of the frustrated antiferromagnetic Heisenberg
quantum spin chain, there is no ordered zero-temperature
phase.\cite{Mermin 66} However, the quantum spin chain at zero
temperature may be mapped to a classical spin chain at finite
temperature. From the non-linear sigma model\cite{Chakravarty 88}
it is known that at least for the unfrustrated Heisenberg model
the temperature $T$ of the classical chain is linked to the quantum spin $S$ by
\begin{equation}
T\propto 1/S \,.
\end{equation}
The classical spin chain at finite temperatures is disordered due
to the Mermin-Wagner theorem;\cite{Mermin 66} but it is ordered
at $T=0$. Reconsidering the arguments at the beginning of this section,
replacing temperature by inverse spin, one sees that the $T=0$ classical
spin chain provides the required commensurate and incommensurate 
ordered low-temperature phases.

In the case of the relationship between the AKLT model and the
antiferromagnetic Heisenberg chain, one has to consider a general
bilinear-biquadratic spin chain as the simplest generalized
Hamiltonian containing both models; it actually shows a phase
transition between an ordered commensurate and an ordered
incommensurate phase in the classical limit. The disorder point
in that case, which was identified as the AKLT model,\cite{Scholl
96a} is thus a quantum remnant of the classical phase
transition.

For a classical frustrated antiferromagnetic Heisenberg chain,
there is a similar phase transition. For $\alpha < \alpha_{c} =
0.25$, the chain is antiferromagnetically ordered:
\begin{equation}
\langle {\mathbf S}_{0}{\mathbf S}_{x}\rangle \propto \cos qx
\end{equation}
with $q=\pi$. For $\alpha > \alpha_{c}$, there is spiral
order 
\begin{equation}
\langle {\mathbf S}_{0}{\mathbf S}_{x}\rangle \propto \cos q(\alpha)x
\end{equation}
with 
\begin{equation}
q(\alpha) = \arccos (-1/4\alpha) .
\end{equation}
In analogy to the bilinear-biquadratic spin chain, we may
therefore predict the presence of a disorder point for a certain
$\alpha_{D}$, exhibiting the same properties as listed
above. Indeed, we can identify an $\alpha_{D}=0.284(1)$,
which meets the above criteria to numerical precision.
Correlations become incommensurate in real space at this point
(see Fig.\ \ref{fig:real_disorder}). The wave number $q$ for
$\alpha > \alpha_{D}$ (Fig.\ \ref{fig:q_disorder})
obtained from fits to a two-dimensional Ornstein-Zernike
correlation function
\begin{equation}
\langle {\mathbf S}_{0}{\mathbf S}_{x}\rangle \propto \cos q(\alpha)x \frac{\exp
  (-x/\xi(\alpha))}{\sqrt{x}}
\end{equation}
shows the expected singular behavior; the singularity is roughly
square-root-like. Figure \ref{fig:correl_disorder} shows a
minimum of $\xi\approx 1.20$ at this point, and a very steep
slope of $\xi$ for $\alpha<\alpha_{D}$. It is numerically
not infinite but clearly much bigger than the slope on the
incommensurate side. We are not able to show the dimensional
reduction at the disorder point numerically; at very small
correlation lengths, it is hard to distinguish between purely
exponential and mixed exponential-power law behavior. At the
disorder point, it seems easier to obtain a purely exponential fit
to the correlation function; but the data does not seem precise enough
that we would want to make a definite statement here.

In the
case of the AKLT disorder point this identification was easily possible
due to the analytically known properties of its ground state. As
all other data ties in extremely well into a very plausible
physical scenario, we are convinced that our identification is
correct.

For $0.37 < \alpha < 0.375$ we find the associated Lifshitz
point, where the $S(q)$ structure function develops a two-peak
structure (see Fig.\ \ref{fig:sq_disorder}). There is no
particular behavior of the structure function at the disorder
point apart from a maximal broadening due to the minimum in
$\xi$. This point has already been found by Pati et al.\cite{Pati
96a,Pati 96b} to be at 0.39(1). As we have investigated longer
chains at higher precision, we consider our result to be more
precise. In any case, this minor disagreement has no direct
physical implications. 

The particular feature of the Lifshitz point 
is the development of a {\em doubly-degenerate} structure of the excitation
spectrum. Let us remark that the existence of two fundamentally
different spectra has already been predicted by Allen and
S\'{e}n\'{e}chal;\cite{Allen 95} they had numerical data suggesting
that spectra switch at $\alpha\approx 0.4$. 

Let us consider closed chains (all statements in the
following are for chains of {\em even} length). In the AKLT
phase, a closed chain can be simulated numerically by adding a spin
$\frac{1}{2}$ at each chain end, {\em as long as the energy to
excite the bonds of these spins to the chain exceeds the bulk gap
energy}. We find numerically for this modified chain:
\begin{itemize}
\item $\alpha<\alpha_{L}$: The
ground state is given by the lowest $0-$ state. The first bulk
excitation is given by the lowest $1+$ state, which does not
degenerate with the lowest $1-$ state.
\item $\alpha>\alpha_{L}$: At the Lifshitz point, the
lowest $1+$ state degenerates with the lowest $1-$ state, which
was higher in energy in the single-state spectrum. This
corresponds to the double peak structure evolving in $S(q)$:
classically speaking, the two degenerate states correspond to
spin waves $\cos qx$ (even parity) and $\sin qx$ (odd parity).
The ground state is still given by the lowest $0-$ state.
\end{itemize}
These observations can be reproduced in unmodified open chains,
if edge excitations are excluded from the spectrum (see Figs.\
\ref{fig:mag50} and \ref{fig:spectra}, discussion in Section V).

Let us briefly discuss the behavior of the gap and the string
order parameter. As can be seen from figures \ref{fig:gap} and
\ref{fig:sop}, the maximum of neither is associated with one of the
two special points just discussed. This is no surprise in the
case of the gap, which has no particular relationship to the
disorder point phenomenon. The maximum of the string order
parameter lies at $0.33 < \alpha < 0.335$, clearly separated from
the disorder point. In our previous study\cite{Scholl 96a} we
found the maximum of the string order parameter to be at the
disorder point. This was however a particular feature due to the
identification of the AKLT point as the disorder point in the
bilinear-biquadratic $S=1$ spin chain. As the string order
parameter is particularly adapted to the AKLT model, it showed
its maximum there. In our present study, the disorder point need
not be (and obviously is not) associated with the frustrated
Hamiltonian ``closest'' to the AKLT model in a generalized
coupling space.

\subsection{The Next-Nearest Neighbor AKLT Phase} 

In the limiting case $\alpha=\infty$, the frustrated chain
decomposes into two unfrustrated chains on the even and odd
sites. Each of these chains can be adequately described by the
conventional AKLT model. We thus use the next-nearest neighbor AKLT
model as shown in Fig.\ \ref{fig:aklt} as starting point for
our argumentation. Observe that in an open chain, there are two
free $S=\frac{1}{2}$ spins at each chain end, which we link up by
nearest-neighbor singlet bonds. There are therefore no free end
spins, and the ground state of an open chain is not degenerate.
This can be verified numerically. The low-lying bulk excitation
spectrum retains its doubly-degenerate structure.
The ground state energy per site should asymptotically behave as
$E_0(0)\alpha$, where $E_0(0)$ is the ground state energy of the
unfrustrated chain. Fig.\ \ref{fig:ground} shows that the asymptotic
behavior is already reached for intermediate $\alpha$, lending support
to our variational ansatz.
We expect an excitation gap of
$\Delta(\alpha)=\Delta(0) \alpha$ 
in the $\alpha\rightarrow\infty$ limit. 
Our numerical calculations (Fig.\ \ref{fig:gap}) show that the asymptotic
behavior of the gap is already approached for intermediate $\alpha$, lending
further support to the variational ansatz. The observed gap exceeds the
asymptotically expected gap, as it costs more energy to excite a chain which
is still (weakly) linked to the other one. 
 
In the limit of very strong frustration, we expect also a
correlation length which is twice the unfrustrated correlation
length, due to the doubling of the lattice spacing:
$\xi(\alpha\rightarrow\infty)=12.04$.
For the correlation
length we observe numerically a drop away from the transition point to a
plateau of $\xi\approx 10$ at $\alpha \approx 2$, with an increase of $\xi$
for larger $\alpha$. The correlation length is smaller than asymptotically
expected, corresponding to the too large gap. 
We cannot make any statement on the large-$\alpha$ behavior, as the
DMRG precision becomes insufficient. 

Let us now address the question whether there is an order
parameter characterizing this phase in analogy to the
conventional string order parameter in the AKLT phase.
Considering the next-nearest neighbor AKLT model, the natural
generalization of the string order parameter is given by
\begin{equation}
{\cal G}_{\pi}^{z}(i,j) = \langle S^{z}_{i} ( \exp
\sum_{k=i+2,\ldots}^{k\leq j}i\pi
S^{z}_{k})S^{z}_{j} \rangle ,
\end{equation}
where $i$ and $j$ are both even (odd), i.e.\ on the same
subchain. At least for $\alpha=\infty$, this must be
a good order parameter.

We observe numerically 
that 
${\cal G}$ vanishes in the AKLT phase with a decay length
substantially shorter than the spin-spin correlation length. In
the NNN-AKLT phase it does not exhibit a finite value for
$|i-j|\rightarrow\infty$ (consider Fig.\ \ref{fig:sop12}). The
decay can be very well fitted to an exponential; a power-law
decay is excluded for the values we have considered. In contrast
to the conventional string order parameter, our generalization is
thus not an order parameter. However, it {\em does} characterize
the nature of this phase in accordance with our analytical model:
above $\alpha\approx 1$, the decay lengths are typically 
much longer than the associated
spin-spin correlation lengths: for $\alpha=2$, the ratio is
already of the order of 10. 
We argue that the difference to the
Haldane phase is given by the restoration of the $Z_{2}\times
Z_{2}$ symmetry on the chain, as characterized by the
disappearance of the conventional string order parameter. In the
AKLT picture this is graphically represented by the two
nearest-neighbor singlet bonds at the chain ends. In the finite
frustration case, we characterize the symmetry on each subchain
as ``almost'' broken: obviously it is broken on the isolated
subchains; but the coupling between the subchains (weaker with
increasing $\alpha$) restores the symmetry on a length scale much
longer than the system correlation length. The following simple
picture can help to illustrate this phenomenon physically: the
difference between our AKLT phase and the exact AKLT state is that
in the AKLT phase there exist bound pairs of solitons in the
hidden (string) order, and the same applies to the subchains in
our NNN-AKLT phase, but now there is a nonzero probability of
having a bound pair with solitons sitting on different subchains,
which destroys the long-range string order inside subchains on
the scale which is roughly the mean distance between soliton
pairs. 
For the Heisenberg point, variational studies \cite{Yamamoto}
 indicate that this mean distance is about 60 lattice sites.
However,
we have no argument at the moment concerning the $\alpha$ dependence 
of this length scale. Further work is necessary
to fully understand this phenomenon.
 
\subsection{The First Order Phase Transition at $\alpha=0.7444(6)$}

The remaining question is how the change from the AKLT to the
NNN-AKLT phase at $\alpha=0.7444(6)$ 
can be characterized. Basically, we have to decide
between (i) no transition, but a gradual change; (ii) a
first-order phase transition; (iii) a continuous phase
transition. Let us recall that in the related
bilinear-biquadratic $S=1$ quantum spin chain there is a
continuous phase transition on the incommensurate side of the
disorder point at the Lai-Sutherland point.\cite{Sutherland 75,Fath 93}
In the following, we
want to discuss our numerical and analytical evidence which
definitely excludes a critical point and thus a continuous phase
transition, and clearly indicates a first-order transition.
Let us first present the raw data.

In the AKLT phase below the transition,
we observe a {\em finite} correlation length peaking at the transition
with a value of $\xi \approx 18$ (see Fig.\ \ref{fig:correl}). 
This data, obtained from chains of $L\leq 380$,
clearly excludes a continuous phase transition, which would require
a divergent correlation length on {\em both} sides of the transition. 
In the NNN-AKLT phase, the correlation length is much longer.
{\em Immediately above} 
the transition, our numerical data does not allow the extraction of a
reasonable correlation length. We can thus not decide whether it
is finite or divergent at the transition in the NNN-AKLT phase. 
This behavior corresponds to a
pronounced
peak in the structure function, but there is no particular
behavior of the peak location $q$.
Apart from the exclusion of a continuous phase transition, the apparent
discontinuity in the correlation length strongly suggests a first order
transition.

We observe a finite gap $\Delta(\alpha)$ (Fig.\ \ref{fig:gap})
everywhere in the same range. This fact is obscured by the
presence of parasitic low-lying states corresponding to edge
excitations, as explained in Section V. The minimal gap is rather
small, $\Delta\approx 0.10$, to be compared with a variational
prediction of $\Delta=0.325$. At the transition, the precision of
the gap is not too high; estimates we give are at the lower
bound. In the AKLT phase, the observation of a finite gap ties in
with the observation of a finite correlation length. The case of
the NNN-AKLT phase we will discuss below.

Our main argument in favor of the first order
transition is the clearly discontinuous disappearance of the
string order parameter (Fig.\ \ref{fig:sop}). We
observe numerically a jump of $0.085$ (20 percent of its maximum value) 
between $\alpha=0.74375$ and
$\alpha=0.74500$. Up to $\alpha=0.74375$, the string order parameter decays
almost linearly; at this point the slope increases about sixtyfold.
We think it is therefore extremely unlikely
that there is a crossover from this linear behavior to an extremely
strong power-law decay (as in a continuous transition), but identify this
behavior as a discontinuous jump.

A first-order transition would be most neatly identified by a
discontinuous derivative of the ground state energy per spin.
Numerically, we find it very difficult to clearly identify such a
discontinuity. Though the correlation length is finite at least in the
AKLT phase, it is long
enough to suggest a rather soft first-order transition. The presence
of degenerate edge states at the transition further obscures the
numerical data.

Another characteristic feature of a first-order transition is the so-called
{\em level crossing}: at the transition, the energy of one of
the two ground states involved drops below the other one. Typically, certain
wave function symmetries change at the transition, and close to the 
transition, in each phase there should be a trace of the ground state of
the other phase. Considering the non-local order parameter, one may not
expect that the symmetry changed is revealed in a change of parity or
the total spin of the ground state, two quantities we control. Actually,
the ground state is $0+$ in both phases. But its degeneracy
changes from 4 (AKLT) to 1 (NNN-AKLT), indicating the change of some more
complex
wave function symmetry. 

We propose the following physical scenario at the transition, which is
strongly supported by our numerical data, and which gives a mechanism for
the first-order transition: 
\begin{itemize}
\item {\em Below} the transition, in the AKLT phase, we find a $1+$ edge
excitation (see Section V, Figs.\ \ref{fig:mag50} and \ref{fig:spectra}),
which degenerates with the ground state at the transition. This edge
excitation is a {\em precursor state} of the transition: In the $0+$,
$1-$ and $1+$ state we observe that the center of
the chain is characterized by an exponential decay of correlations. We
say it is in the {\em bulk phase}, which is just the AKLT phase.
Close to the transition the chain ends, 
however, belong to a different {\em edge phase}: 
as we calculate
spin-spin correlations symmetrized around the center, this is evidenced
by a clear change in the spin-spin correlation function.
For longer chains at fixed $\alpha$, 
we find that the bulk region grows, but not the edge region,
showing that we are in fact dealing with an end effect.   
We call this a {\em pseudo coexistence} of phases: 
though they coexist on the finite chain, the edge phase is {\em not}
extensive; in the thermodynamical limit it is just a boundary effect,
so there is no true coexistence. We may however consider the chain
ends as a {\em nucleation center} for the new phase: we suppose this
is because the open chain ends allow a lowering of energy 
by replacing an AKLT chain with free end spins by two chains, whose free
end spins can be bound by singlets, which lower the energy.
We identify the edge phase with the NNN-AKLT phase,
because its influence exists only close to the transition and the
string order parameter does not develop a finite long-range value in
the edge phase.
At the transition the AKLT bulk phase, whose correlation
length is finite all the way, is pushed out entirely, 
as the NNN-AKLT edge phase
becomes extensive. At this point,
the $0+$, $1+$ and $1-$ states are degenerate (to numerical
precision). The bulk excitation of the
AKLT phase has a finite gap up to the transition.
\item {\em Above} the transition, in the NNN-AKLT phase, there is no
bulk-edge separation in the ground state. There is a low-lying $1\pm$ pair
of states (almost) 
degenerate with the ground state at the transition, emerging
as low-lying excitations. 
The magnetization is concentrated in the
chain ends, shifting towards the center, as the energy cost of this
excitation approaches that of a NNN-AKLT bulk excitation. 
Calculating increasing chain lengths for fixed $\alpha$,
the magnetization remains at the chain ends.
These excitations must thus be classified as edge excitations,
until the gap between $1\pm$ and $0+$ becomes of the order of the
bulk gap between $2\pm$ and $1\pm$.
We observe the following interesting
phenomenon: Up 
to the discontinuous jump, the string order parameters as calculated
in the $0+$ and the $1\pm$ states agree to numerical
precision, excluding the edge regions. 
After the transition, the string order decays fast to zero in the
$0+$ ground state, but remains
non-zero in the bulk of the $1\pm$ states much longer before decaying.
With increasing $\alpha$ (as the excitation wanders into
the bulk) the decay behaviour aproaches that of the $0+$ state, and the
string order starts fluctuating strongly. The effect disappears at 
$\alpha\approx 0.80$.
The correlation
length observed in the center region ties in well 
with the AKLT correlation length
just below the transition.

We suggest, on the above grounds, that those $1\pm$ states are the
trace of the old AKLT ground state in the new phase, however,
``polluted" by parasitic edge excitations. Then, the described
spectrum behavior is consistent with the level crossing picture
and with all the other data indicating the first order
transition. Assuming that the spin wave velocity does not diverge
at the transition, our gap curve would suggest that the
correlation length is finite on the NNN-AKLT side of the
transition also.
\end{itemize}

We are therefore led to locate a first-order phase transition at
$\alpha_{T}=0.7444(6)$, in very good agreement with the
naive analytical prediction $\alpha_{T}^{var}=0.75$ (see next section).

In the following we discuss in more detail our variational and
numerical approaches.

\section{Analytical Results: Variational Approach to the
Frustrated Spin Chain}

In the following we are presenting our variational calculations.
Such calculations are very useful to understand the nature of
the ground state and the excitations of quantum systems, and
even get a quantitative estimate of energies.
However, 
strictly speaking, no variational result can be considered a
strong argument, as far as the nature or very existence of
a phase transition is concerned. Nevertheless, in many
cases a variational study of the ground state {\em and} the
elementary excitations, while looking for the possible points
where the gap closes, turns out to be useful and gives
an important hint of the actual system behavior. For example,
variational studies of the solitonic excitations in the $S=1$ chain
\cite{NeuMik96,Mik92,Gomez89,Mik-chaos95}
allow one to reproduce qualitatively the
structure of the phase diagram in the presence of anisotropies,
biquadratic exchange, and an external magnetic field; a variational
study of the Shastry-Sutherland-type\cite{ShastrySuth81} solitons in the
dimer order qualitatively captures the picture of the transition
from the dimerized to the nondimerized phase.\cite{AKunpub}
Therefore we
will present here the variational results for frustrated S=1
chain as arguments complementing the numeric findings. The good
agreement between the physical assumptions underlying the variational
ansatz and the numerical results shows that we capture essential parts
of the physics of the frustrated $S=1$ chain. 

\subsection{Ground State and Elementary Excitations: A
Naive Approach}

In the most naive way, one can attempt to describe the
ground state of the frustrated $S=1$ chain as being the AKLT
state below an $\alpha_T^{var}$ and NNN-AKLT state above. 
The ground state energy per spin $E_0$ then would be
$E^{AKLT}_0= -4/3 +4\alpha/9$ in the AKLT phase and
$E^{NNN}_0=-4\alpha/3$ in the NNN-AKLT phase, which gives a rough
(but surprisingly good, see Fig. \ref{fig:ground}) estimate 
for the transition point $\alpha_T^{var}=\frac{3}{4}$. 

The elementary excitations in
the AKLT phase then can be studied in a soliton approach in
the
spirit of Refs. \onlinecite{FathSolyom93,NeuMik96}. Technically this 
is most easily
done in the matrix product states formalism.\cite{Fannes+89,Klumper+91-93}
Let us briefly summarize the results.
The AKLT state can be represented in the form of a trace over the
matrix product
\begin{equation}
|AKLT\rangle =\mbox{Tr} (\prod_{i=1}^N g_i)\,,
\label{MP_aklt}
\end{equation}
where
\begin{equation}
g_i={1\over \sqrt{3}} \left(
\begin{array}{rr}
|0\rangle_i & -\sqrt{2} |+\rangle_i \\
\sqrt{2} |-\rangle_i & -|0\rangle_i
\end{array}\right)
\label{aklt_matrix}
\end{equation}
is a 2$\times$2 matrix composed of the spin states
of the $i$-th site. The soliton (``crackion,'' in the
terminology of Fath and Solyom) state $|C_n^\mu\rangle$,
describing the soliton in the string order located at the $n$-th
site and having
$S^z=\mu$, $\mu=0,\pm1$, can be written as
\begin{equation}
|C_n^\mu\rangle =\mbox{Tr} (\prod_{i=1}^{n-1} g_i (\sigma^\mu g_n)
\prod_{i=n+1}^{N} g_i )\,,
\label{cr_AKLT}
\end{equation}
where $\sigma^\mu$ denotes the Pauli matrices in the spherical
basis. Physical excitations with a definite momentum can be easily
constructed as $|C^\mu(k)\rangle =\sum_n e^{ikn}
|C_n^\mu\rangle$, and their dispersion is
\[
\varepsilon^\mu(k)=
{\langle  C^\mu(k)| (\widehat H -E_0) |C^\mu(k)\rangle
\over \langle  C^\mu(k)| C^\mu(k)\rangle}\;.
\]

The averages can be calculated using the transfer matrix
technique (see Refs.\ \onlinecite{Klumper+91-93,TotsukaSuzuki95}), 
and finally,
after a simple but lengthy calculation, one arrives at the following
formula for the dispersion law of the soliton excitation at
$\alpha < 0.75$:
\begin{eqnarray}
\varepsilon^\mu(k)&=&{14\over 9} +{26\over27} \alpha
+{160\alpha-18 \over 27}\,\cos(k) -{14\over9}\alpha\cos(2k)
\nonumber \\
&+&(2-26\alpha/3){3+5\cos(k) \over 5+3\cos(k)}
\label{disp_AKLT}
\end{eqnarray}
Because of the isotropy of the problem, all three branches with
different $\mu$ are degenerate.

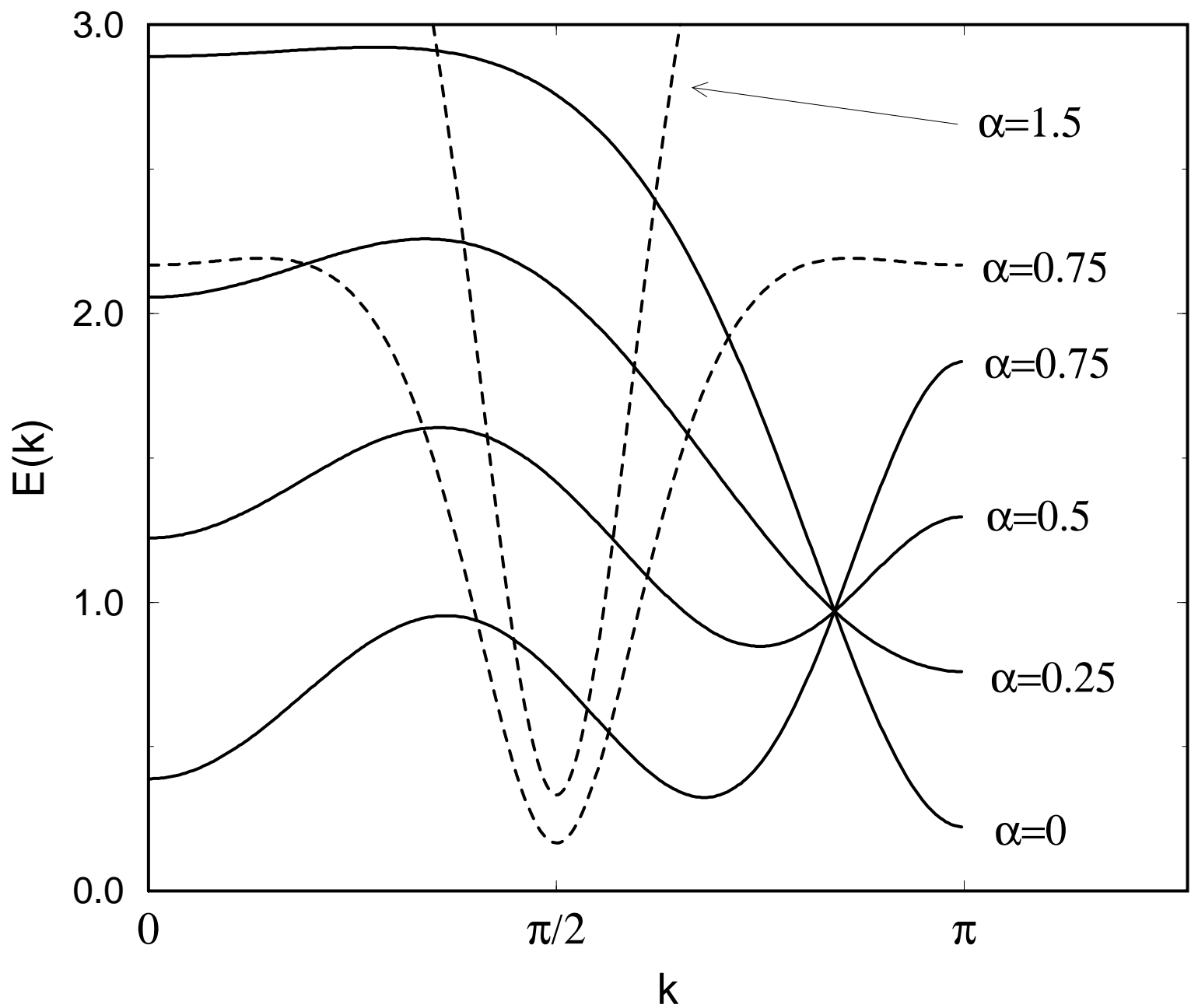
\begin{figure}
\caption{\label{fig:disp} Typical dispersion curves of a soliton
(``crackion'') excitation above the AKLT (solid lines) and
NNN-AKLT (dashed lines) states, for different values of the
next-nearest neighbor coupling $\alpha$.}
\end{figure}

A few typical dispersion curves at different values of $\alpha$
are displayed in Fig. \ref{fig:disp}. One can see that above some
critical $\alpha=\alpha_L^{var}\simeq0.32$, the minimum of the
excitation energy is found for a momentum $k=q_0\not=\pi$, and
$q_0$ tends to $2\pi/3$ as $\alpha$ tends to
$\alpha_T^{var}=\frac{3}{4}$ (see Fig. \ref{fig:qpeak}). One may
speculate this point $\alpha_L^{var}$ can be identified with the
Lifshitz point, though in fact there are no incommensurate
correlations in the AKLT state. Note that in the numerical
calculations the lowest excitation becomes doubly degenerate when
$\alpha$ crosses the Lifshitz point. This can be easily
explained\cite{Allen 95} by the fact that the two minima of the
dispersion curve at $k=\pm q_0$ are physically inequivalent if
$q_0\not= 0,\pi$.

The gap  does not disappear at the transition point
($\Delta^{var}(\alpha=0.75)\simeq0.325$), indicating a first-order phase
transition (or absence of a phase transition). The $\alpha$
dependencies of the gap $\Delta^{var}$ and of the 
wavevector $q_0^{var}$ with
minimal excitation energy
obtained from this simple calculation qualitatively agree with
the numerical data (see Figs. \ref{fig:gap} and
\ref{fig:qpeak}).

On the other side of the transition point we take the NNN-AKLT
state (Fig.\ \ref{fig:aklt})
as a variational ground state, and the soliton dispersion
in the thermodynamic limit $L\to\infty$ (when the chains become
decomposed in the NNN-AKLT picture) can be obtained from
(\ref{disp_AKLT}) by first setting $\alpha$ to zero, replacing $k\to
2k$, and finally scaling the whole expression by $\alpha$; this gives
for $\alpha > 0.75$ the gap $E_g=2\alpha/9$ (similarly to
$E_g=2/9$ at the Heisenberg point $\alpha=0$) and $q_0=\pi/2$ as
wavevector with minimal excitation energy. Actually, as it
follows from the numerics, $E_g(\alpha)\to \alpha E_g(0)$ only
asymptotically at $\alpha\to\infty$,  but one can also see from
the numerical data that this asymptotic linear behavior above
$\alpha=0.75$ sets in rather quickly.

\subsection{$\mathbf{4\times4}$ Matrix Product Variational Ansatz}

The ``purely AKLT'' description presented above is of course not
very satisfactory: in fact, we did not even have any variational
parameters and simply compared two VBS configurations being
intuitively good candidates for the ground state. However, it is
easy to see that those two choices are not the only possible
ones: for example, at $\alpha=0.75$ the energy per spin of the 
completely dimerized valence bond state is exactly the same as that of the
AKLT and NNN-AKLT configurations. One thus may try to consider
some more general variational wavefunction capable of
interpolating between different VBS states.

We have found that such
wavefunctions can be constructed in the matrix product states
formalism, at the price of going to a higher matrix dimension
and considering larger clusters.
One possible simplest choice for the elementary matrix is
\begin{eqnarray}
\Gamma_{12}&=&\sum_{ij}| t_{1i}t_{2j} \rangle \left\{ A\delta_{ij}
(1\!\!1 \otimes 1\!\!1) + iB\varepsilon_{ijk}(\sigma_k
\otimes 1\!\!1) \right.\nonumber\\
&+&\left. iC(\sigma_i\otimes\sigma_j) \right\}\;,
\label{4by4}
\end{eqnarray}
where the matrix state $\Gamma_{12}$ lives on a cell consisting
of two adjacent spins 1 and 2. $\sigma_i$, $i=x,y,z$ are the
usual Pauli matrices (in the cartesian basis), $1\!\!1$ denotes
the 2$\times$2 identity matrix, and $| t_i\rangle$ denotes the
triplet of spin-1 states in the cartesian basis:
\begin{eqnarray*}
&&| t_x \rangle =-(1/\sqrt{2})(|+\rangle -|-\rangle)\,,\\
&&| t_y \rangle =(i/\sqrt{2})(|+\rangle +|-\rangle)\,, \\
&&| t_z \rangle =|0\rangle \,.
\end{eqnarray*}
Since both the Pauli matrices and the triplet wavefunctions
$|t_i\rangle$ behave as vectors under rotations,
the matrix (\ref{4by4}) behaves as a scalar. Therefore,
the matrix product state
\begin{equation}
|\Omega\rangle =\mbox{Tr}\,(\prod_l \Gamma_{2l-1,2l})\,,
\label{wf}
\end{equation}
constructed from such elementary matrices,
obeys the rotational invariance of the problem
(note that the usual AKLT matrix
(\ref{aklt_matrix}) is unitary equivalent to $(1/\sqrt{3})\sum
\sigma_i |t_i\rangle$, and this is the only possible
rotationally invariant ansatz if the dimension is 2$\times$2 and
the matrix lives at one site, but in higher dimensions and for
larger cluster sizes the number of choices rapidly increases).

The wavefunction (\ref{wf}) has the remarkable property that it
interpolates smoothly between the AKLT state ($A=B=1/3$, $C=0$),
the completely dimerized state ($A=1/\sqrt{3}$, $B=C=0$), and the
NNN-AKLT state ($A=B=0$, $C=1/3$).

The quantum averages can be calculated in the usual way;
however, the complexity of solving the analytic eigenvalue
problem for the 16$\times$16 transfer matrix $G=\Gamma_{12}^*
\Gamma_{12}$ forced us to restrict ourselves to the case of
real coefficients $A$, $B$, $C$. Then, setting the largest
eigenvalue of $G$ to 1, one obtains
the normalization condition
\begin{equation}
3(A^2+2B^2+3C^2)=1\,
\label{nrm}
\end{equation}
which leaves us with two independent real variational parameters.
The variational expression for the ground state energy per spin
\begin{eqnarray}
E_{var}&=& -4\alpha/3 + 4B^2(4\alpha-3)
     -3(A^2-B^2) \\
\label{gs44}
&+&2(A-B)\Big\{2A^3 +10A^2B +AB^2-2A/3\nonumber\\
&+&5B^3 -10B/3 +\alpha \big[-10A^3 -10A^2B \nonumber \\
&&\quad -2AB^2 +16(A+B)/3 -2B^3\big]\Big\}\,\nonumber
\end{eqnarray}
can be minimized numerically; the resulting dependence of the
ground state energy on $\alpha$ is presented in Fig.\
\ref{fig:ground}.
The main feature is that though the discontinuity at the
transition is less distinct than in the ``naive'' picture,
and the transition point shifted towards larger $\alpha\simeq
0.81$,
the transition still is found to be first order.
At the Heisenberg point $\alpha=0$ the variational result for 
the ground state energy is $E_{var}^{\text{min}}=-1.364$, being
slightly better than the AKLT value $-{4\over 3}$. However, the
disadvantage of the ansatz (\ref{wf}) is that it explicitly
breaks the translational invariance, and thus the ground state
has a built-in dimerization which is always nonzero.

The lowest excitations above this variational ground state can be
calculated using the single-mode approximation in the spirit of Arovas
et al. \cite{ArovasAH88} Since we have now two spins involved in 
the elementary matrix (\ref{4by4}), 
it is natural to write down the wavefunction
of the excited state in the form
$|\widetilde{C}^\mu(k)\rangle=\sum_n e^{ikn}
|\widetilde{C}_n^\mu\rangle$, where
\begin{eqnarray}
&& |\widetilde{C}_n^\mu\rangle =\mbox{Tr}
(\prod_{l=1}^{n-1} \Gamma_{2l-1,2l} \cdot
\Gamma_C^\mu \cdot
\prod_{l=n+1}^{N} \Gamma_{2l-1,2l} )\,, \\
\label{mod_exc}
&& \Gamma_C^\mu=  \big(\widehat{S}^\mu_{2n-1} +
\lambda(k)\cdot\widehat{S}^\mu_{2n}\big) \Gamma_{2n-1,2n}\,. \nonumber
\end{eqnarray}
Here $\widehat{S}^\mu_l$ denote the components of the spin-1
operator at the $l$-th site, 
 $\lambda$ is an additional (complex) variational parameter whose
value has to be determined numerically for each value of the
wavevector $k$, and $k$ now varies from $0$ to $\pi/2$ since the
elementary cell is doubled. The resulting gap dependence on
$\alpha$ is shown in Fig. \ref{fig:gap}. One can see that there
is a local minimum around $\alpha=0.75$, but the gap
in the transition region is considerably overestimated, even
comparing to the naive approach described in the previous
subsection; we attribute this fact to the built-in breaking of
the translational invariance in the ansatz (\ref{wf}) as
explained above.

\section{Numerical Results: Density Matrix Renormalization Group
Calculations}

\subsection{General Remarks}

Using the Density Matrix Renormalization Group (DMRG), we have
calculated for this problem (i) the gap between the ground state
and the lowest excitation, (ii) the magnetization for the lowest
excited state, (iii) the string order parameter ${\cal O}$ and
the string correlator ${\cal G}$ and (iv) the spin-spin
correlation function. For details on the DMRG we refer to the
literature.\cite{White 92,White 93b} The DMRG is particularly
suited for the problem under study as it is not limited to small
systems like exact diagonalization and as it is not plagued by
the negative sign problem which Quantum Monte Carlo typically
encounters in frustrated systems at very low temperatures.

For our calculations, we have studied chains of a length up to
$L=380$ and typically kept $M=250$ block states in each iteration. To
reduce both memory usage and improve program execution speed, we
have used the following DMRG features:
\begin{itemize}
\item We have used both the total magnetization and left-right
parity as good quantum numbers. This reduces storage, but also
thins out the Hilbert space, giving faster convergence of the
implemented exact diagonalization, and allows for fast
classification of the spectrum.
\item We have implemented a prediction algorithm which gives a
guess for the eigenstates of a DMRG step based on the eigenstates
of the preceding DMRG step. This algorithm,\cite{Scholl 96z}
similar in spirit to the one introduced recently by 
White,\cite{White 96,White 96b} allows
for a substantial reduction of the number of iterations needed in
the exact diagonalization, truncating it by a factor of up to 10.
It should however be mentioned that the speed-up due to the
prediction algorithm is biggest when the studied system has a
rather short correlation length; so its use is somewhat limited
there where most performance would be needed. A further problem
is that it does not cut the time needed for the calculation of
expectation values, a dominant feature of our calculations.
\end{itemize}

It is important to realize that (unlike in exact diagonalization
studies) it is not sufficient to extrapolate results to the
thermodynamic limit in $L$ only. The performance of the DMRG
depends crucially on the number $M$ of block states kept; as a
rule of thumb, the number of states $M$ to be kept increases
dramatically close to critical points or phase transitions, which
reflects the increasing number of low-energy fluctuations or 
competing states. The
precision of the DMRG is indicated by the truncation error, which
allows for extrapolations to the exact $M=\infty$ result. Let us
remark that good agreement between DMRG and exact diagonalization
results for a given $M$ is not necessarily an indicator of good
DMRG precision: exactly diagonalizable systems are notoriously small,
and our results indicate that DMRG errors build up severely with
system length.

We found that close to the disorder and Lifshitz points (a
region where the ground state should have a relatively simple
valence bond structure) $M=80$ is sufficient to give highly
precise results; near the phase transition, convergence up to
$M=200$ was poor, forcing us to go up to $M=250$ states. It was
not possible to go to the limit of very high $\alpha$: in this
case, the chain essentially decouples into two subchains, leading
to a ground state which can be understood as a product state of
the ground states of two unfrustrated chains. The description of such a
product state implies a dramatic increase in $M$, as already
observed in other works.\cite{White 96,Scholl 96b}

The numerical investigation of the phase transition was further
complicated by the fact that the DMRG works best for open
boundary conditions. At the transition point it was therefore not
possible to use periodic boundary conditions, as the precision
obtained for open ones was already only moderate. Open boundary
conditions may however introduce additional edge states into the
spectrum which suggest often radically different physical
properties. A good example is provided by the unfrustrated 
open integer spin
chain with spin $S$, where there are two effectively free $S/2$
spins at each chain end. For the $S=1$ chain these free spins
introduce the well-known Kennedy triplet, which degenerates with
the ground state. One therefore has the fifth state as the first
bulk excitation. In the frustrated chain, 
the situation will be shown to be
not always as clear. To identify the lowest bulk excitation, it
is therefore necessary to calculate $\langle S^{z}_{i}\rangle$
for all low-lying states. Edge excitations due to the open chain
ends can be identified by very small $\langle S^{z}_{i}\rangle$
in the chain center and big $\langle S^{z}_{i}\rangle$ at the
chain ends and thus excluded.

\subsection{Calculation of Correlations} 

For $\alpha<\alpha_D$ {\em spin-spin correlation lengths} can be
obtained by a fit of the spin-spin correlations to a law
$(-1)^{x}\exp (-x/\xi)/\sqrt{x}$, which is in all cases
extremely well obeyed, except exactly at the disorder point. 
Note that all DMRG correlation lengths 
are underestimations of the true correlation length. For
the longest correlation length in that region (at $\alpha=0$), we
obtain $\xi\approx 5.8$, underestimating the true result by about
3 percent. As the correlation length decreases as well as the
truncation error with $\alpha$ (it is essentially 0 at the
disorder point), the error diminishes. The correlation length at
the disorder point we estimate to be precise to the order of 1
percent or better. All other correlation lengths lie in between. For
$\alpha>\alpha_D$, correlation lengths are obtained by
a least square fit of the data to a law $\cos qx \exp
(-x/\xi)/\sqrt{x}$, with $q$ and $\xi$ to be determined.
Extremely good such fits could be obtained. Alternatively, one
may simply plot $|\langle {\mathbf S}_i{\mathbf S}_j\rangle|\sqrt{|i-j|}$
logarithmically. As the logarithm is not very sensitive close to
correlation function maxima, simply drawing an upper
straight-line envelope gives a very simple estimate which is
hardly worse despite the incommensurate correlations. Again, the
correlations obtained in that way are underestimated. Considering
the results for various $M$ we claim that we underestimate at
worst by about 20 percent around the transition; at the
transition itself, the data is inconclusive
on the NNN-AKLT side. Just below the transition,
edge effects become strong, and have to be excluded from the calculation
of the correlation length. In short chains, the bulk behaviour 
may not be visible. Edge effects can be identified by calculating longer
chains: the region of bulk behavior extends with $L$.

The {\em string order parameter} is a quantity
particularly suited for treatment by the DMRG: It reaches its
thermodynamical limit very fast; the decay to the thermodynamical
limit is on a scale of the order of half the bulk spin-spin
correlation length, as was already observed for unfrustrated spin
chains.\cite{White 93b} Its convergence to its exact value in $M$
is also very fast, unlike the convergence of the correlation
length.  

\subsection{Gap Calculations: The Spectrum of Open Chains} 

The calculation of the {\em bulk} excitation gap was the most
difficult calculation performed, because of the already mentioned
problem of edge excitations inherent to open chains as used by
the DMRG. Ground state and lowest excited state
energies were first extrapolated in $M$ for
fixed $L$ (using the roughly linear dependence of the error in energies
on the DMRG truncation error) and then extrapolated in $L$ using quadratic
convergence laws, which are very well verified. This
allows us to obtain gaps with a precision up to $10^{-3}$, at
least $0.02$ at the transition. 
The essential difficulty arises from the {\em correct 
identification of the lowest bulk excitation}. For most values
of frustration, this can be nicely done by
calculating the $\langle S^{z}_{i}\rangle$ distribution along the chain
for  $S^{z}_{total}=S_{total}$ (see Fig.\ \ref{fig:mag50}). Typically, it
is very easy to distinguish true bulk excitations from edge excitations.
To devise a more stringent criterion, it is useful to study the rather
rich behavior of the low-energy excitation spectrum of {\em open} frustrated
chains, as the DMRG deals best with open systems. 
In fact, though the open chain spectrum is more complicated, the mechanism
of the phase transition is better revealed here, as the
associated symmetry breaking is obvious in the presence or
absence of effectively free end spins. We find the following
scenario (all statements for chains of even length; example spectra
are given in Fig.\ \ref{fig:spectra}; the arrow indicates the gap
energy):

\begin{itemize}
\item $\alpha<\alpha_{D}$: The ground state is given by
the lowest $0+$ state: there is an odd number of singlet bonds in
the bulk, the two effectively free end spins are linked by an
extremely weak singlet bond, giving a total even number of
singlet bonds. Exciting this weak singlet bond gives a $1-$
triplet excitation, degenerating with the ground state (the Kennedy
triplet). The first bulk excitation is given by the $2+$
quintuplet excitation, combining a bulk and an edge excitation.
The $2-$ quintuplet is not degenerate with the $2+$ quintuplet,
and can be identified as an edge excitation.

\item $\alpha_{D}<\alpha<\alpha_{L}$: The open boundary conditions introduce
{\em a further parasitic edge excitation} $1+$: this
edge excitation we identify from its evolution with $\alpha$ 
as a precursor of the phase transition and the NNN-AKLT phase. 
Even parity coupling of edge spins
is energetically disfavored in the AKLT phase, $1+$ lies much
higher than $1-$.

\item $\alpha_{L}<\alpha<\alpha_{T}$: As already
described for the closed chain, the {\em bulk} excitations
degenerate in pairs of odd and even parity excitations with
identical total spin. The ground state is still the same $0+$
state, there is a degenerate Kennedy triplet $1-$. The lowest
bulk excitations are now given by the degenerate $2\pm$
quintuplet excitation, combining a $1\pm$ bulk excitation with a
$1-$ edge excitation (see Fig.\ \ref{fig:mag50}). The $1+$ edge
excitation lowers its energy with $\alpha$, degenerating with
the ground state at $\alpha_{T}$.
For numerical
calculations it is important to realize that there may be {\em two}
$S=1$ excitations below the true bulk excitation. Following the gap
calculation
procedure described by Pati et al.,\cite{Pati 96b} we can explain
the difference between their and our gap curve for $\alpha\approx
0.5$ arguing that they have measured the energy difference
between the two edge excitations. The numbers we obtain that way
agree perfectly with theirs. As the $1+$ edge
excitation degenerates with the $1-$ excitation at the transition, 
a vanishing gap is suggested, as reported in their work. This is
however {\em not} the true bulk excitation.

\item $\alpha=\alpha_{T}$: For the behavior of the spectrum at the transition,
we refer to Section III D.

\item $\alpha>\alpha_{T}$: Beyond the transition, the ground state $0+$ is
unique, as the Kennedy triplet disappears. The
situation is basically not very complicated for the lowest bulk
excitation: there is a doubly degenerate $1\pm$ bulk excitation as below
the transition.
The actual location of this state in the complete spectrum of the open
chain
however varies: {\em just above} the transition, there are
low-lying $1\pm$ edge excitations. The bulk excitation is hidden
in the lowest $2\pm$ state. 
For {\em intermediate and
strong frustration}, the chain decouples effectively into two
subchains which are only weakly interacting: The nearest-neighbor
singlet bonds at the chain ends 
become increasingly easy to excite for increasing $\alpha$, giving four
free spins-$\frac{1}{2}$. These couple into 16 states
degenerating for $\alpha\rightarrow\infty$, coupling into one $0+$,
two $1-$, one $1+$ and one $2+$ state. The bulk gap, on the other hand,
scales with $\alpha$. 
As soon as edge
excitation energies drop below the bulk gap energy, the bulk excitation 
will only be present in higher spin states, which have to be 
identified by the magnetisation. 
Numerically, one can keep the bulk excitation in the $1\pm$ state by
increasing the interaction strengths at the chain ends, to disfavor
edge excitations.

\end{itemize}

We have calculated the lowest energy-states
in the following sectors of the Hilbert space: $0+$, $1+$, $1-$, $2+$
and $2-$, to verify the above scenario by considering the bulk
magnetization. For sufficiently long chains, it was always possible to
clearly separate bulk from edge excitations as in the above scenario.

\section{Conclusion}

The numerical results presented above together with the
variational calculations allow to devise a clear and coherent
picture of the behavior of a frustrated $S=1$ isotropic
Heisenberg spin chain. Its behavior is fundamentally governed by
the underlying classical model, which is characterized by a phase
transition from an antiferromagnetic to a spiral ordered phase.
On the other hand, pure quantum effects are prominent. As
predicted from the non-linear sigma model, the system is gapped
for all values of frustration. However, there are two clearly
separated phases present. The first one for small frustrations
can be well understood in terms of the
Affleck-Kennedy-Lieb-Tasaki model and is characterized by a
non-vanishing conventional string order parameter. The classical
phase transition is reflected by the presence of a so-called
disorder point, where the spin-spin correlations become
incommensurate. The classical phase transition is thus not linked
to the first order phase transition found for larger frustration.
The disorder point is clearly related to the disorder point in a
bilinear-biquadratic quantum spin chain, which is exactly the
AKLT model. 
The phase transition found at
$\alpha=0.7444(6)$ is purely quantal in character. It is first order,
characterized by a non-vanishing gap and a finite correlation
length on the AKLT side. We
observe a non-continuous change both in the string order parameter
and in
the correlation length. The $Z_{2}\times Z_{2}$ symmetry broken in the
AKLT phase is thus restored. A string correlator which
considers only every second spin characterizes this phase.
We also suggest that if one includes the
alternation of  $[1+(-1)^i\delta]$ type 
in the nearest-neighbor interaction
in the Hamiltonian (\ref{ham}), there will be a first-order transition
{\em line} in the $(\alpha\delta)$ plane.
Assuming that the transition line is
characterized by vanishing string order, we suggest that it
should be identified with the $(BC)$ line in Fig.\ 3 of Ref.\
\onlinecite{Pati 96a} separating the region with fourfold
degenerate ground state from the region where the ground state is
unique. Since for  $\alpha\to 0$ and large $\delta$ the latter
region coincides with the well-studied dimerized phase,
\cite{KatoTanaka94,Totsuka+95,Yamamoto95} the question arises
whether our NNN-AKLT phase transforms smoothly into the dimerized
phase or there is one more transition line separating the
dimerized and the NNN-AKLT phases. 

\section*{Acknowledgements}

Two of us (U.Sch.\ and A.K.) wish to thank H.-J. Mikeska for his
hospitality at the Institut f\"ur Theoretische Physik, Hannover,
where the collaboration leading to this paper was initiated, and
for fruitful conversations. We thank H.J. Mikeska, 
S. Miyashita, T. Tonegawa,
H. Wagner, and S. Yamamoto for useful discussions. A.K.
acknowledges the financial support by Deutsche
Forschungsgemeinschaft. Numerical calculations were performed
mostly on a PentiumPro 200MHz machine running under Linux.

\begin{figure}
\caption{\label{fig:aklt} Schematic representation 
of the AKLT model and its next-nearest neighbor generalization. 
Circles are
spin-1 sites, dots represent a spin $\frac{1}{2}$, 
and fat links are singlet bonds between spins. Note
the presence of a free spin $\frac{1}{2}$ at each end of the open AKLT chain.
In the NNN-AKLT model, the dashed line represents the underlying chain.}
\end{figure}

\begin{figure}
\caption{\label{fig:real_disorder} Spin-spin 
correlations
for various frustration values just below
and above the disorder point.
We show the logarithm of $|\langle {\mathbf S}_i {\mathbf S}_j \rangle|$
times the square root of the spin-spin distance.
Purely antiferromagnetic correlations can thus be most
easily distinguished from incommensurate ones, which show prominent peaks.}
\end{figure}

\begin{figure}
\caption{\label{fig:correl_disorder} Spin-spin 
correlation length $\xi(\alpha)$ 
in the vicinity of the disorder 
and Lifshitz points. Note that these correlation lengths are systematically 
underestimated by the DMRG. From the known $\alpha=0$ result and the dependence
of the error on the DMRG truncation error we estimate the error to be 3 percent
in the worst case, typically 1 percent or better close to the disorder point.}
\end{figure}

\begin{figure}
\caption{\label{fig:q_disorder} Correlation wave 
numbers $q(\alpha)$ in the vicinity
of the disorder point obtained by fits of the correlation function to
an expected Ornstein-Zernike behavior of the correlations.}
\end{figure}

\begin{figure}
\caption{\label{fig:sq_disorder} 
Structure function $S(q)$ for various values of frustration
$\alpha$. The values of $\alpha$ are, ordered by decreasing $S(\pi)$,
0.3, 0.32, 0.34, 0.36, 0.37, 0.375, 0.38, 0.4, 0.5, 0.6, 0.7, and 1.5. 
Note the
developing double peak structure for
$0.37 < \alpha < 0.375$. The double peak shifts to $q=\pm \pi/4$,
due to the doubling of the lattice spacing in the decoupled chains in the
$\alpha\rightarrow\infty$ limit.}
\end{figure}

\begin{figure}
\caption{\label{fig:sop} String order parameter 
$\lim_{|i-j|\rightarrow\infty} {\cal O}(i,j)$.
The full squares show the string order parameter for the $0+$ ground state;
there is a discontinuous drop of $0.085$ at $\alpha_T$.}
\end{figure}

\begin{figure}
\caption{\label{fig:gap} Bulk excitation gap $\Delta(\alpha)$. 
The precision of the gap values is between 0.001
and 0.02 (close to the transition). Within the error bars, shown values
were chosen to be a lower bound to the exact gap.
For large $\alpha$, the gap approaches
the asymptotic value $\Delta(\alpha)=\alpha\Delta(0)$.
The solid and pointed lines are the analytical results 
for the 2$\times$2 and the 4$\times$4
matrix product ansatz, respectively. 
}
\end{figure}

\begin{figure}
\caption{\label{fig:correl} Spin-spin correlation 
lengths. 
Correlations lengths are systematically underestimated due to the DMRG.
Around the transition the error is maximal, but will generously estimated not
exceed 20 percent.}
\end{figure}

\begin{figure}
\caption{\label{fig:sop12} Decay behavior
of the generalized string correlator for various
frustration values in the nnn-AKLT phase.}
\end{figure}

\begin{figure}
\caption{\label{fig:ground} In this figure variational and
numerical ground state energies are shown. The dashed and solid
lines correspond to the $2\times 2$ and $4\times 4$ variational
ansatz respectively,
the solid squares are numerical results. Numerical errors
are smaller than $10^{-5}$.}
\end{figure}

\begin{figure}
\caption{\label{fig:qpeak}  Dependence of the peak wavevector in
the structure factor on the next-nearest neighbor coupling
constant $\alpha$ is shown. Solid squares are DMRG data,
the solid line is the variational result for the wavevector
with the minimal excitation energy, according to
(\ref{disp_AKLT}).}
\end{figure}

\begin{figure}
\caption{\label{fig:mag50} Magnetization of various
lowest excitations for $\alpha=0.5$. The
$1+$ and $1-$ states are edge excitations; the $1-$ state is the Kennedy  
triplet. The $2\pm$ states are a combination of the $1-$ edge excitation and
a true bulk excitation.}
\end{figure}

\begin{figure}
\caption{\label{fig:spectra} Evolution of the excitation 
spectrum in the AKLT phase. E stands for
pure edge excitation. Note the double degeneracy of the 
first excitation beyond
the Lifshitz point and the appearance of a low-lying edge state beyond the
disorder point. The arrow indicates the states to be compared for 
gap calculations.}
\end{figure}

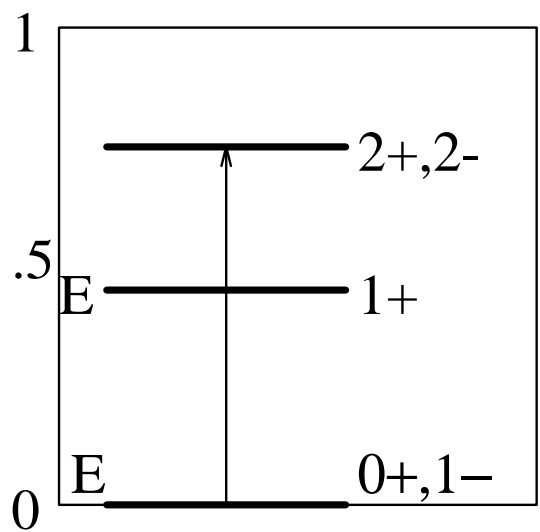
\end{document}